\documentclass[12pt]{article}
\usepackage{authblk}
\usepackage[comma, numbers, square]{natbib}
\usepackage{graphicx}
\usepackage[a4paper,margin=1in]{geometry}
\usepackage{hyperref}
\usepackage{amsmath}
\usepackage{amssymb}
\usepackage{xspace}
\usepackage{color}           

\newcommand{\spacing}[1]{\renewcommand{\baselinestretch}{#1}\large\normalsize}
\spacing{1.25}

\newenvironment{addendum}{%
    \setlength{\parindent}{0in}%
    \small%
    \begin{list}{Acknowledgements}{%
        \setlength{\leftmargin}{0in}%
        \setlength{\listparindent}{0in}%
        \setlength{\labelsep}{0em}%
        \setlength{\labelwidth}{0in}%
        \setlength{\itemsep}{12pt}%
        }
    }
    {\end{list}\normalsize}

\newcommand\ion[2]{#1$\;${\small\rmfamily\rom{#2}}\relax}%
\newcommand*{\rom}[1]{\uppercase\expandafter{\romannumeral #1\relax}}
\newcommand{\kms}{km~s$^{-1}$\xspace} 
\newcommand{\dqe}{DQSL$_\text{east}$\xspace}
\newcommand{\dqw}{DQSL$_\text{west}$\xspace}

\newcommand{\aap}{    {\it Astron. Astrophys.}}

\newcommand{\apj}{    {\it Astrophys. J.}}
\newcommand{\apjl}{   {\it Astrophys. J. Lett.}}

\newcommand{\solphys}{{\it Solar Phys.}}
\newcommand{\ssr}{    {\it Space Sci. Rev.}}

\title{Disintegration of an Eruptive Filament via Interactions with Quasi-Separatrix Layers} 
\date{\vspace{-5ex}}
\author{Rui Liu, Jun Chen, Yuming Wang}
\affil{CAS Key Laboratory of Geospace Environment, Department of Geophysics and Planetary Sciences, University of Science and Technology of China, Hefei 230026, China \\ rliu@ustc.edu.cn}

\begin{document}

\maketitle


\begin{abstract}
The disintegration of solar filaments via mass drainage is a frequently observed phenomenon during a variety of filament activities. It is generally considered that the draining of dense filament material is directed by both gravity and magnetic field, yet the detailed process remains elusive. Here we report on a partial filament eruption during which filament material drains downward to the surface not only along the filament's legs, but to a remote flare ribbon through a fan-out curtain-like structure. It is found that the magnetic configuration is characterized by two conjoining dome-like quasi-sepratrix layers (QSLs). The filament is located underneath one QSL dome, whose footprint apparently bounds the major flare ribbons resulting from the filament eruption, whereas the remote flare ribbon matches well with the other QSL dome's far-side footprint. We suggest that the interaction of the filament with the overlying QSLs results in the splitting and disintegration of the filament. 
\end{abstract}



\section{Introduction}
Solar filaments are cool and dense `clouds' suspended in the hot and tenuous corona. In H$\alpha$, filaments appear in absorption against the disk but in emission above the limb (also termed \emph{prominences} in this occasion, but often used interchangeably with filaments); in EUV, they generally have a dark appearance, but may become brightened from time to time due to heating, e.g., the interface region between the filament and the ambient corona often emits in UV/EUV \cite{Labrosse2010}. Magnetic field lines threading the prominence material, whether they are dipped \cite{ks1957}, helically coiled \cite{kr1974,Low&Hundausen1995}, or flat-topped \cite{Karpen2015}, play a critical role in the formation and suspension of filaments in the corona \cite{Mackay2010}. The interplay between a filament and the field supporting its weight, which involves gravitational settling and radiating cooling, naturally leads to the formation of magnetic discontinuities \cite{Low2015}. Current layers may form at such discontinuities and leave imprints on the surface \cite{Wang2017NC}. 

Filaments are highly dynamic on a wide range of spatio-temporal scales, and it is frequently observed that filament material drains back to the surface during a variety of activities, e.g., in a partial eruption, in which either a significant \cite{Tripathi2009} or minimal \cite{Liu2007} fraction of filament mass is ejected with the surrounding magnetic structure as a coronal mass ejection (CME), or in a failed eruption \cite{Ji2003}, in which the filament initially displays similar dynamic behaviors as a successful eruption but is later confined without being associated with a CME. The filament draining is an important phenomenon, since it might serve as a trigger for filament eruptions \cite{Zhou2006} and can also be used to infer the sign of helicity of filaments \cite{Chen2014}. During the well-studied partial filament eruption on 2011 June 7 \cite{Reale2013,Gilbert2013,Carlyle2014}, falling filament fragments impact on places as far away as one solar radius from the source region: some fragments are significantly affected by inertia and gravity, while others are redirected toward remote magnetic footpoints \cite{vanDriel-Gesztelyi2014}. Wang et al. \cite{WangW2017} reported that a curtain-like structure (CLS) develops during the tornado-like evolution of a quiescent prominence. The CLS consists of myriads of thread-like loops along which heated prominence material slides downward and lands outside the filament channel. The draining results in the disintegration of the prominence and the CLS is suggested to form through magnetic reconnections between the prominence field and the overlying coronal field, with the threads of the CLS representing new magnetic connections channeling filament material. The renowned failed prominence eruption on 2002 May 27 \citep{Ji2003,Alexander2006} also develops a similar CLS after the saturation of the helical kink of the prominence. In a parametric MHD simulation of the same event \cite{Hassanin&Kliem2016}, two distinct magnetic reconnection processes are identified, i.e., the reconnection between the erupting flux and the overlying flux, which leads to the formation of the CLS, and the reconnection in the vertical current sheet between the two legs of the original flux rope, which is responsible for the associated flare.

The impact sites of falling filament material often exhibit EUV brightening. Two physical mechanisms are proposed to explain the brightening \cite{Gilbert2013}:  heating due to the falling material compressing the plasma, or reconnection between the low-lying loops and the magnetic field carried by the impacting material; the former is preferred over the latter for the 2011 June 7 event, as the energy release in emission is much smaller than the kinetic energy of falling material. Here by presenting the observation of a partial filament eruption, we propose a third mechanism, in which the interaction of the eruptive filament with overlying quasi-separatrix layers (QSLs) results in the splitting of the filament and filament material falls toward the footprint of the QSLs, observed as a remote flare ribbon. The UV/EUV brightening at the impact site is mainly due to the field-aligned transport of energy released by magnetic reconnections at the QSLs. This is expected to result in much more intense heating than kinetic impacting, the latter of which carries an energy of $\sim10^{27}$ erg \cite{Gilbert2013} as small as a micro-flare. Magnetic reconnection has been reported to occur between an eruptive filament and its ambient field taking the forms of various structures, e.g., a coronal hole \cite{Zhu2014}, coronal loops \cite{Li2016}, chromospheric fibrils \cite{Xue2016}, or another filament \cite{Zhu2015}. The reconnection between an eruptive filament and its overlying (quasi-)separatrix surfaces is well expected but has not been studied in the literature. In the sections that follow, the instruments and data analysis techniques utilized in this study are briefly introduced in Section 2. The observation is described and analyzed in detail in Section 3. Concluding remarks are given in Section 4. 

\section{Methods}
\subsection{Instruments and Datasets}
The filament eruption occurred in NOAA AR 11936 (S16W32; Figure~\ref{fig:aia}) on 2013 December 31, which is associated with a GOES\footnote{Geostationary Operational Environmental Satellite} M6.4-class flare. According to the GOES 1--8~{\AA} flux, the flare starts at 21:45 UT, peaks at 21:58 UT, and by 22:20 UT the flux has decayed to a level halfway between the maximum and the pre-flare background. The filament eruption and the flare are imaged by the Atmospheric Imaging Assembly (AIA \cite{lemen12}) onboard the Solar Dynamics Observatory (SDO \cite{pesnell12}). AIA is equipped with seven EUV and two UV narrow-band channels spanning a broad temperature range and being sensitive to different heights and structures in the solar atmosphere. In this study, we focused on three passbands, i.e., 131~{\AA} (\ion{Fe}{21} for flare, peak response temperature $\log T = 7.05$; \ion{Fe}{8} for active regions, $\log T = 5.6$ \cite{ODwyer2010}) in which flaring plasma in the corona is well characterized, 304~{\AA} (\ion{He}{2}, $\log T = 4.7$) which features the chromosphere and transition region, and 1600~{\AA} which features both the upper photosphere (continuum) and transition region (\ion{C}{4}, $\log T = 5.0$). The instrument takes full-disk images with a spatial scale of 0.6 arcsec pixel$^{-1}$ and a cadence of 12 s for EUV and 24 s for UV passbands. 

The eruption results in a CME observed in white light by the C2 and C3 coronagraphs of the Large Angle and Spectrometric Coronagraph Experiment (LASCO \cite{Brueckner1995}) onboard the Solar and Heliospheric Observatory (SOHO). The CME spans an angular width of about $\sim\,$90 deg and propagates at an average speed of $\sim\,$300 \kms, but does not exhibit a coherent, bubble-like structure (see SOHO LASCO CME Catalog\footnote{\url{https://cdaw.gsfc.nasa.gov/CME_list/UNIVERSAL/2013_12/univ2013_12.html}}), as seen in a typical CME. The CME does not exhibit a bright core either, suggesting that only an insignificant fraction of the filament mass escapes with the CME.    

The magnetic environment in which the eruption occurs is monitored by photospheric magnetograms taken by the Helioseismic and Magnetic Imager (HMI \cite{hoeksema14}) onboard SDO. The vector magnetograms for active regions used in this study are disambiguated and deprojected to the heliographic coordinates with a Lambert (cylindrical equal area; CEA) projection method, resulting in a pixel scale of $0.03^\circ$ (or 0.36 Mm) \cite{Bobra2014}.

\subsection{Field Extrapolation}
To understand the magnetic connectivities in AR 11396, we extrapolated the coronal potential field from the photospheric boundary with the Fourier transformation method \cite{Alissandrakis1981}. Though a zero-order approximation of the real coronal field, potential field maintains the basic topology, due to the robustness of structural skeletons of magnetic field \cite{Titov2007,Titov2009}, which has been demonstrated by earlier studies employing various coronal field models \cite{demoulin06,demoulin07,Liu2014}, and also by switching on and off a ``pre-processing'' procedure on the photospheric boundary \cite{Liu2016SR}. In this study, the pre-processing procedure \cite{wiegelmann06} is applied to a pre-flare vector magnetogram (Figure~\ref{fig:topology}a) to best suit the force-free condition. The vector magnetogram is taken from the Space-Weather HMI Active Region Patches (\texttt{hmi.sharp\_cea}) series and $2\times2$ rebinned. The calculation was carried out within a box of $640\times 272\times 272$ uniformly spaced grid points, corresponding to $466\times198\times198$ Mm. The photospheric magnetic flux is roughly balanced within the field of view (Figure~\ref{fig:topology}a), with the ratio between positive and (absolute) negative flux being 1.1. 

We also extrapolated a nonlinear force-free field (NLFFF), using the code package developed by T.~Wiegelmann \cite{wiegelmann04,wiegelmann12} and taking the same processed vector magnetogram as the extrapolation boundary, but failed to identify a coherent magnetic flux rope, which would otherwise be spotted as a volume of enhanced twist number ($|\mathcal{T}_w| \ge 1 $ \cite{Liu2016}) enclosed by quasi-separatrix layers (QSLs), which is quantified by squashing factor as explained below.

\subsection{Squashing Factor}
Squashing factor $Q$ \cite{Titov2002} measures the spatial rate of change in magnetic connectivities. For a mapping through the two footpoints of a field line $\Pi_{12}:  \mathbf{r}_1(x_1,\ y_1)\mapsto \mathbf{r}_2(x_2,\ y_2)$, the squashing factor $Q$ associated with the field line is \cite{Titov2002}
\begin{equation} \label{eq:q}
Q\equiv \frac{a^2+b^2+c^2+d^2}{|\mathrm{det}\,D|},
\end{equation}
where $a,b,c,d$ are elements of the Jacobian matrix 
\begin{equation} \label{eq:d12} 
D=\left[\frac{\partial \mathbf{r}_2}{\partial \mathbf{r}_1}\right]= 
\begin{pmatrix}
\displaystyle \frac{\partial x_2}{\partial x_1} & \displaystyle \frac{\partial x_2}{\partial y_1} \\
\displaystyle \frac{\partial y_2}{\partial x_1} & \displaystyle \frac{\partial y_2}{\partial y_1}
\end{pmatrix} \equiv 
\begin{pmatrix}
a & b \\
c & d
\end{pmatrix}.
\end{equation}
Quasi-separatrix layers (QSLs) are defined as high-Q structures, where the field-line mapping has a steep yet finite gradient, whereas $Q\rightarrow\infty$ at topological structures \cite{Titov2002}. The visualization of these complex three-dimensional structures can be facilitated by calculating $Q$ in a 3D volume box. This is done by stacking up $Q$-maps in uniformly spaced cutting planes \cite{Liu2016}.  

\subsection{Decay Index}
How the magnetic field decays with increasing height plays an important role in regulating the behavior of solar eruptions \cite{Torok2007}. Theoretically, a toroidal flux ring is unstable to expansion if the external poloidal field $B_{\mathrm{ex}}$ decreases sufficiently rapidly as the height increases, typically when the decay index $n=-d\ln B_{\mathrm{ex}}/d\ln h$ exceeds 3/2 \cite{Kliem&Torok2006}. As one cannot clearly distinguish $B_{\mathrm{ex}}$ from the flux-rope field, we follow the usual practice, i.e., to approximate $B_{\mathrm{ex}}$ with a potential field \cite{WangD2017}. In our calculation, $n=-d\ln B_t/d\ln h$, where $B_t=\sqrt{B_x^2+B_y^2}$ denotes the transverse component of the extrapolated potential field. Precisely speaking, the field component orthogonal to the axial current of a flux rope provides the downward $\mathbf{J}\times\mathbf{B}$ force, but $B_t$ also serves as a good approximation since potential field is typically orthogonal to the PIL, along which the flux rope in equilibrium resides. 

\section{Results}
\subsection{UV/EUV Observation}
\begin{figure} 
	\centering
	\includegraphics[width=\hsize]{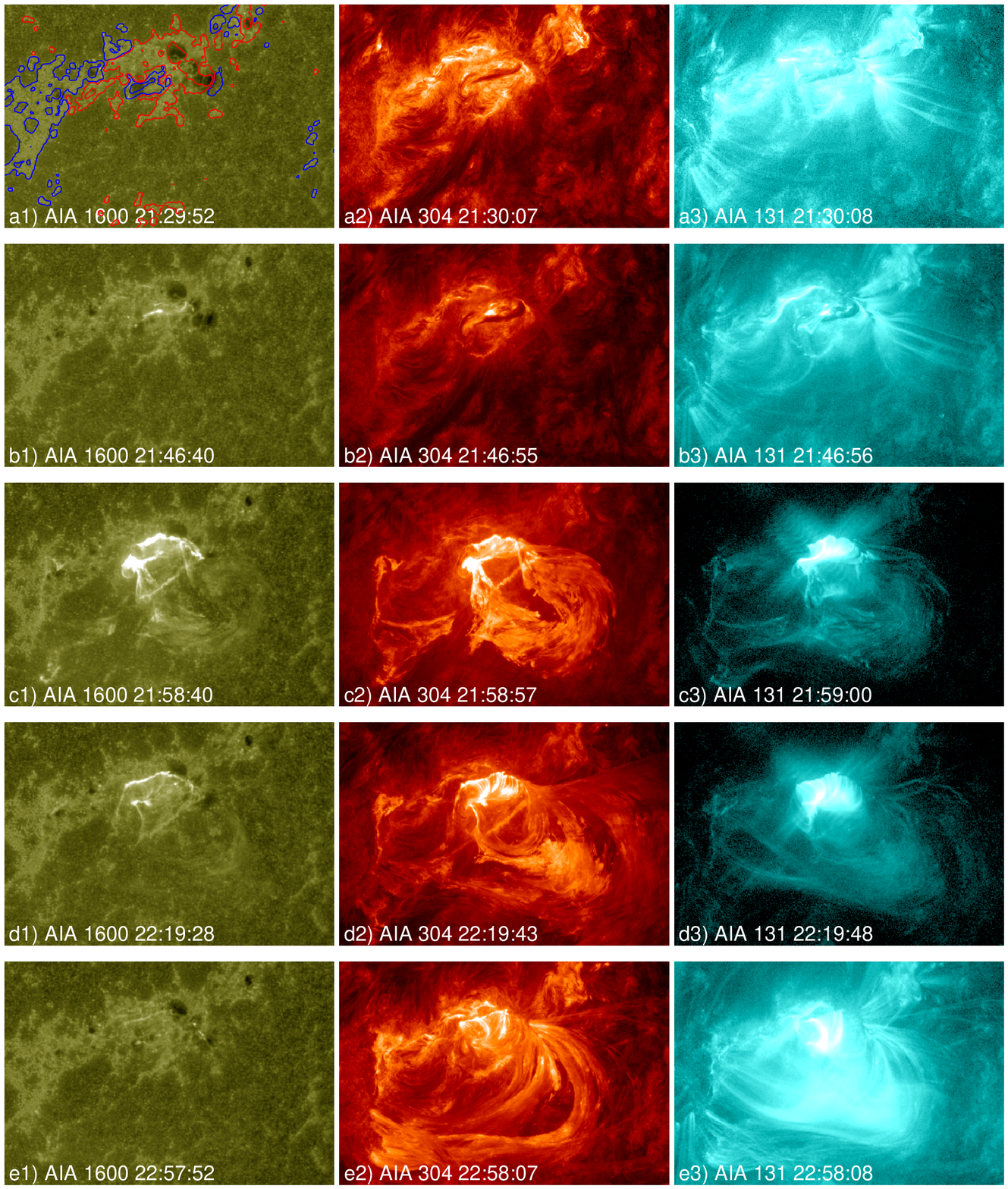}
	\caption{Snapshots of the filament eruption observed by SDO/AIA. From left to right column the images in 1600, 304, and 131~{\AA} passband are shown in a logarithmic scale. The field of view is $370''\times 250''$. Panel (a1) is superimposed by the contours of the line-of-sight component of the photospheric magnetic field. The contour levels are set at $\pm50$ and $\pm500$ Gauss, with blue (red) indicates negative (positive) polarities. An animation of AIA images is available at \url{http://staff.ustc.edu.cn/~rliu/preprint/disintegration.avi}. \label{fig:aia}}
\end{figure}

The filament is located along the major polarity inversion line (PIL) separating the positive-polarity flux concentration in the north and the negative-polarity flux concentration in the south (see also Figure~\ref{fig:topology}(a)). It is noteworthy that some weak positive-polarity fluxes are scattered around the negative flux concentration, hence producing a $\delta$-spot configuration, which is known to be prolific in flare productions. The filament has an eastern hook curving southward and a western hook slightly curving northward, so that it takes a revers-S sigmoidal shape (Figure~\ref{fig:aia}(a2)). Starting from 21:30 UT, the filament rises slowly. At 21:46 UT, a bright point appears below the filament, which is associated with the triggering of the filament eruption (Figure~\ref{fig:aia}(b2)). The eruption results in two flare ribbons with intense brightening in 1600~{\AA} at two sides of the major PIL in the core region (Figure~\ref{fig:aia}(c1); labeled R1 and R2 in Figure~\ref{fig:topology}(b). These two ribbons correspond to the footpoints of the post-flare arcade observed in 131~{\AA} (Figure~\ref{fig:aia}(c3, d3)). The northern ribbon, however, extends and half circles around the southern ribbon, although this extension (labeled R1e in Figure~\ref{fig:topology}(b)) is relatively weak. There is a third ribbon (labeled R3 in Figure~\ref{fig:topology}(b)), which is even weaker and located remotely in the plages of negative polarity in the east. R3 takes a north-south orientation, which is temporarily visible at 1600~{\AA} during the flare main phase but its visibility at 304 and 131~{\AA} lasts into the decay phase (Figure~\ref{fig:aia}(b2--e2 and b3--e3)). It is noteworthy that R1e extends further southeastward and connects with R3. 

As the filament rises upward rapidly, one can see that filament material drains back to the surface, either along the filament legs back to its feet or towards the V-shaped connection between R1e and to R3. The latter produces a curtain-like structure (Figure~\ref{fig:aia}(e2 and e3)), spreading out along the remote flare ribbon. At 22:19 UT, the eruptive filament seems to split into two branches. The lower branch still anchors at the original place, i.e., the two ends of the post-flare arcade (Figure~\ref{fig:aia}(d2)); the higher branch shares the western foot with the lower one, but its eastern foot seems to anchor at the V-shaped connection between R1e and R3. Meanwhile, field connectivity between R3 and R1e is established, as evidenced by the high-lying hot loops in 131~{\AA} (Figure~\ref{fig:aia}(d3, e3)) above the post-flare arcade connecting R1 and R2. By 23:30 UT the filament is completely disintegrated, with the bulk of filament mass draining back to the surface, whereas a small fraction is supposedly ejected into interplanetary space with the CME. 

\subsection{Magnetic Configuration}
From the EUV observations, it is clear that the draining process is not a free falling but shaped by magnetic field. It is remarkable that except to the original filament feet, filament material mainly drains back to the remote ribbon R3. One can see from Figure~\ref{fig:aia}(a1 and a2) that R3, located within a negative-polarity region in the east, is separated by a positive-polarity region from the filament along the major PIL in the center of AR 11936. This suggests that for the filament material to fall to R3, magnetic reconnection must occur to join these two topologically distinct regions, which is evidenced by the high-lying hot loops connecting R3 and R1e in the wake of the filament disintegration (Figure~\ref{fig:aia}(d3, e3)). To understand the role of magnetic topology in this process, we performed the extrapolation of the coronal potential field based on the photospheric Bz component of a vector magnetogram taken immediately prior to the flare (Figure~\ref{fig:topology}(a)), and further calculated the map of squashing factor $Q$ in this field (Figure~\ref{fig:topology}(c)).

\begin{figure} 
	\centering
	\includegraphics[width=\textwidth]{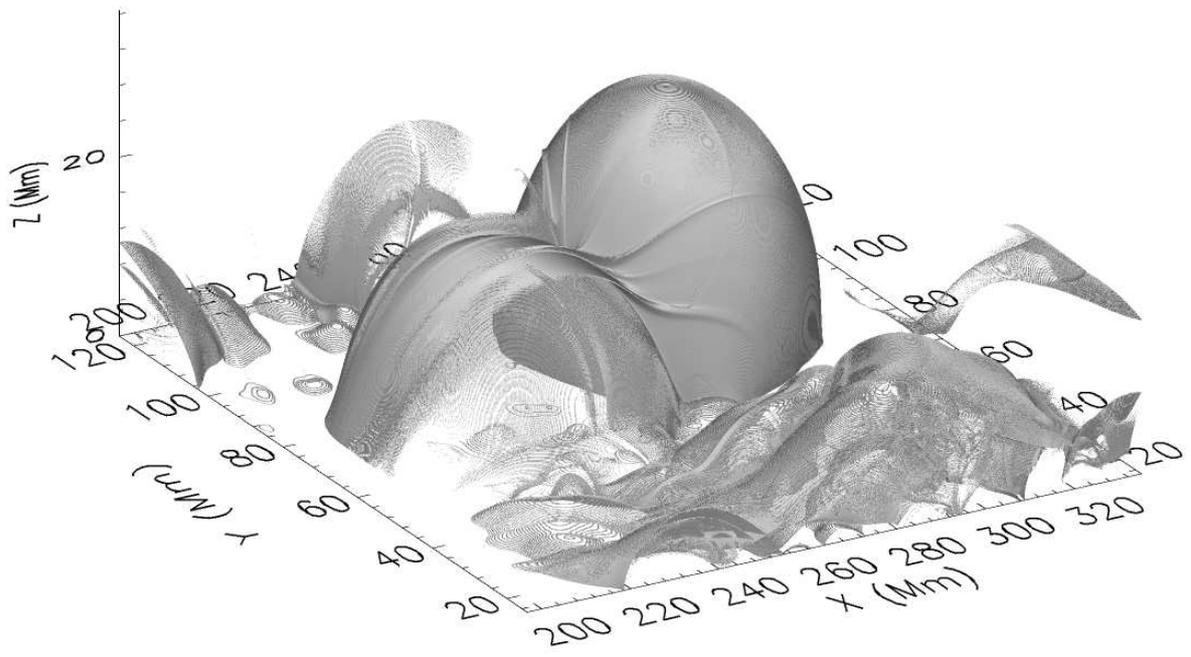}
	\caption{Isosurface of $\log_{10}Q=4.5$ shown in a 3D perspective \label{fig:qsl}}
\end{figure}

The three-dimensional distribution of $Q$ features two dome-like QSLs (DQSL hereafter) conjoining together (Figure~\ref{fig:qsl}), whose footprint on the surface can be seen in Figure~\ref{fig:topology}(c). However, the eastern dome (\dqe hereafter) has smaller $Q$ values than its western counterpart (\dqw hereafter), especially for its northern and southern facets. As a result, the former's footprint in the north and south is less well defined as the latter's, which is characterized by an oval-like high-Q line. In comparison to an AIA 1600~{\AA} image taken during the flare main phase, which is deprojected using the same CEA method as the HMI vector magnetogram (Figure~\ref{fig:topology}(b)), one can see that the ribbon R1 and its half circular extension R1e match well with the footprint of the western dome, that R2 corresponds to the east-west oriented high-Q lines inside the western dome, and that the remote ribbon R3 compares favorably with the eastern section of \dqe's footprint. This topology is different from that in typical circular-ribbon flares, in which a single dome-like (quasi-)separatrix surface may account for the circular ribbon \cite{Liu2015,Zhang2015,Yang2015}. Here, the two conjoining DQSLs constitute a hyperbolic flux tube (HFT) \cite{Titov2002}, which is considered to be a preferential location for the concentration of strong currents and the subsequent rapid dissipation \cite{Titov2003,Galsgaard2003,Aulanier2005}.

We then sampled the segment of PIL that is located in between R1 and R2, the footpoints of the major post-flare arcade, by randomly clicking along the PIL to pick up some representative points (marked by colored stars in Figure~\ref{fig:topology}(a and c)), and then calculate decay index $n$ at different heights at these selected points. In Figure~\ref{fig:decay} we plot the profile of $n$ as a function of $Z$ at the selected points, with the same color code as in Figure~\ref{fig:topology}(a and c). One can see that the $n(Z)$ profile tends to have a saddle-like shape for points located closer to the intersection of the two QSL domes, but $n$ increases monotonically with $Z$ for points far away from this intersection. All the profiles converge at heights beyond about 90 Mm. For a saddle-like profile, the eruptive structure may become unstable at a lower height than in a case with a monotonic profile, but the deep saddle bottom may provide an additional constraint \cite{WangD2017}. This is consistent with the decay index in an $XZ$ cutting plane at $Y=98.3$ Mm (indicated by the horizontal dashed line in Figure~\ref{fig:topology}a), which displays a `bump' ($X\simeq270$ Mm) of small decay indexes ($n<1$) up till $Z\simeq50$ Mm in the center (see also Figure~\ref{fig:decay}c) and a `dip' of large decay indexes ($n\ge1$) to the west of the bump down to $Z\simeq5$ Mm, suggesting that a structure located in the west would be more susceptible to the torus instability.  In comparison to the heights of both DQSLs (Figure~\ref{fig:qsl}), one can see that the filament, especially its eastern section, would start to interact with the QSLs, before it becomes torus unstable. 

\section{Conclusion \& Discussion}
\begin{figure} 
	\centering
	\includegraphics[width=\hsize]{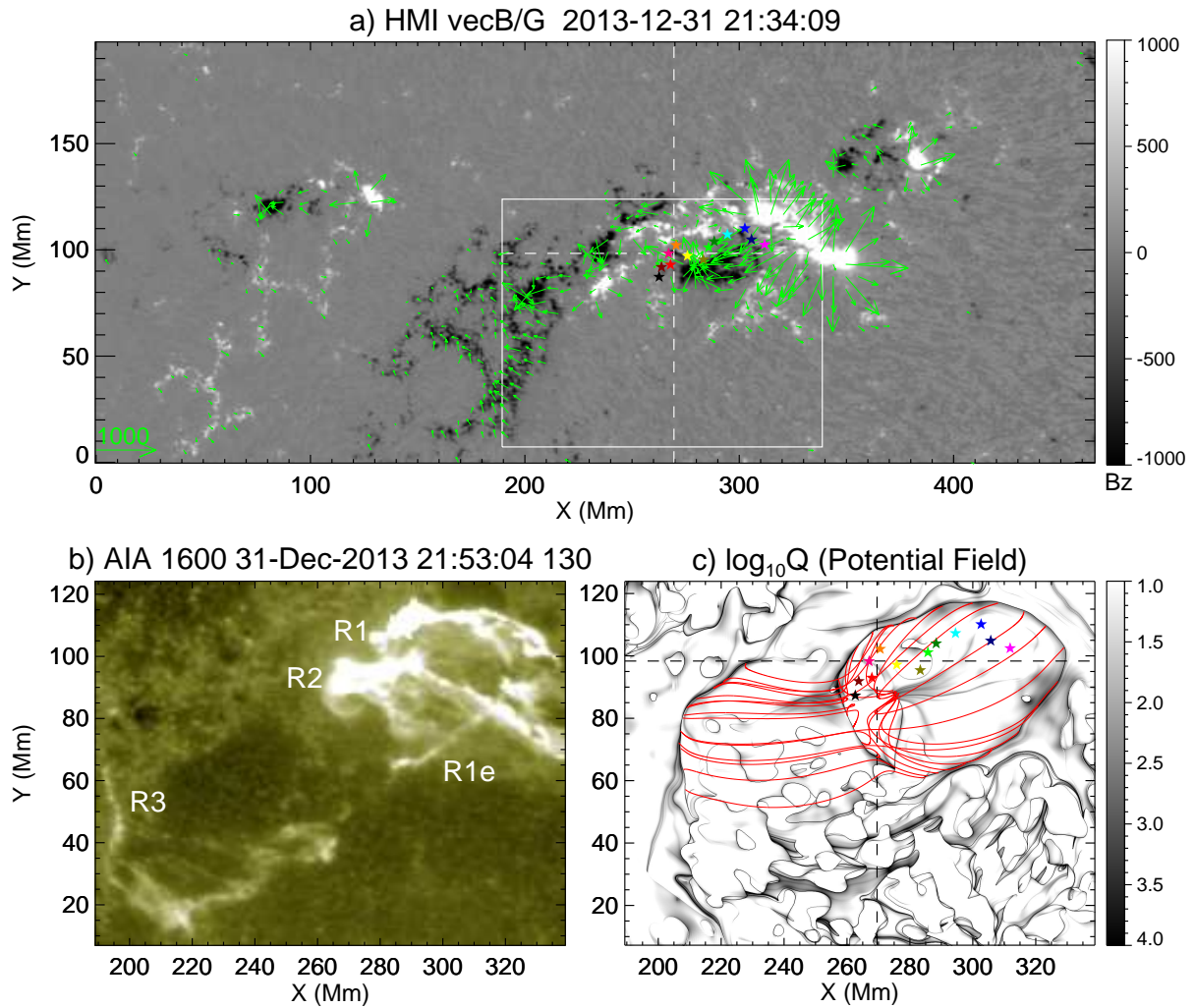}
	\caption{Magnetic configuration of AR 11936. (a) Vector magnetogram taken immediately prior to the filament eruption on 2013 December 31. White and black colors refer to the positive and negative normal ($B_z$) field component, respectively, which are saturated at $\pm1000$ Gauss (see the color bar). Green arrows represent the tangential field component, with the arrow length at 1000 Gauss indicated at the lower left corner. The rectangle denotes the field of view shown in Panels (b) and (c). The two dashed lines indicate the positions of the $XZ$ and $YZ$ cutting planes in Figure~\ref{fig:decay}. Panel (b) shows an AIA 1600~{{\AA}} image during the rising phase of the flare. The two major flare ribbons, labeled R1 and R2, correspond to the footpoints of the post-flare arcade. The remote ribbon is labeled R3. The half-circular extension of R1 is labeled R1e, which extends further southeastward to connect with R3. The AIA image is projected with the same CEA method as the vector magnetogram in (a).  Panel (c) shows the photospheric $Q$-map, superimposed by field lines (red) traced from the high-Q lines. Colored stars in (a) and (c) indicate the selected points along the PIL to calculate the decay index as a function of $Z$ in Figure~\ref{fig:decay}(a).  \label{fig:topology}}
\end{figure}

Based on the above analysis, we conclude that the disintegration of the eruptive filament is due to its interaction with two conjoining DQSLs, because the filament material drains back not only to the filament's original footpoints, but to the remote flare ribbon R3 which matches well with the eastern section of \dqe's footprint. Tracing field lines from both DQSLs' footprints demonstrate that there exist field connectivities between the eastern section of \dqe's footprint and \dqw's circular-shaped footprint (Figure~\ref{fig:topology}c). Thus, although the filament is located underneath \dqw, its interaction with \dqw would naturally result in filament material moving along magnetic field towards R3. We suggest that this is achieved by three dimensional magnetic reconnection at the DQSLs, where current layers form due to the rising filament exerting pressure upon the DQSLs. The reconnection results in the restructuring of the filament, as evidenced by the splitting of the filament and the redirection of filament mass towards the remote flare ribbon. This scenario is also supported by the observation of the high-lying EUV hot loops above the post-flare arcade, in the wake of the disintegration of the filament. 

\begin{figure} 
	\centering
	\includegraphics[width=\hsize]{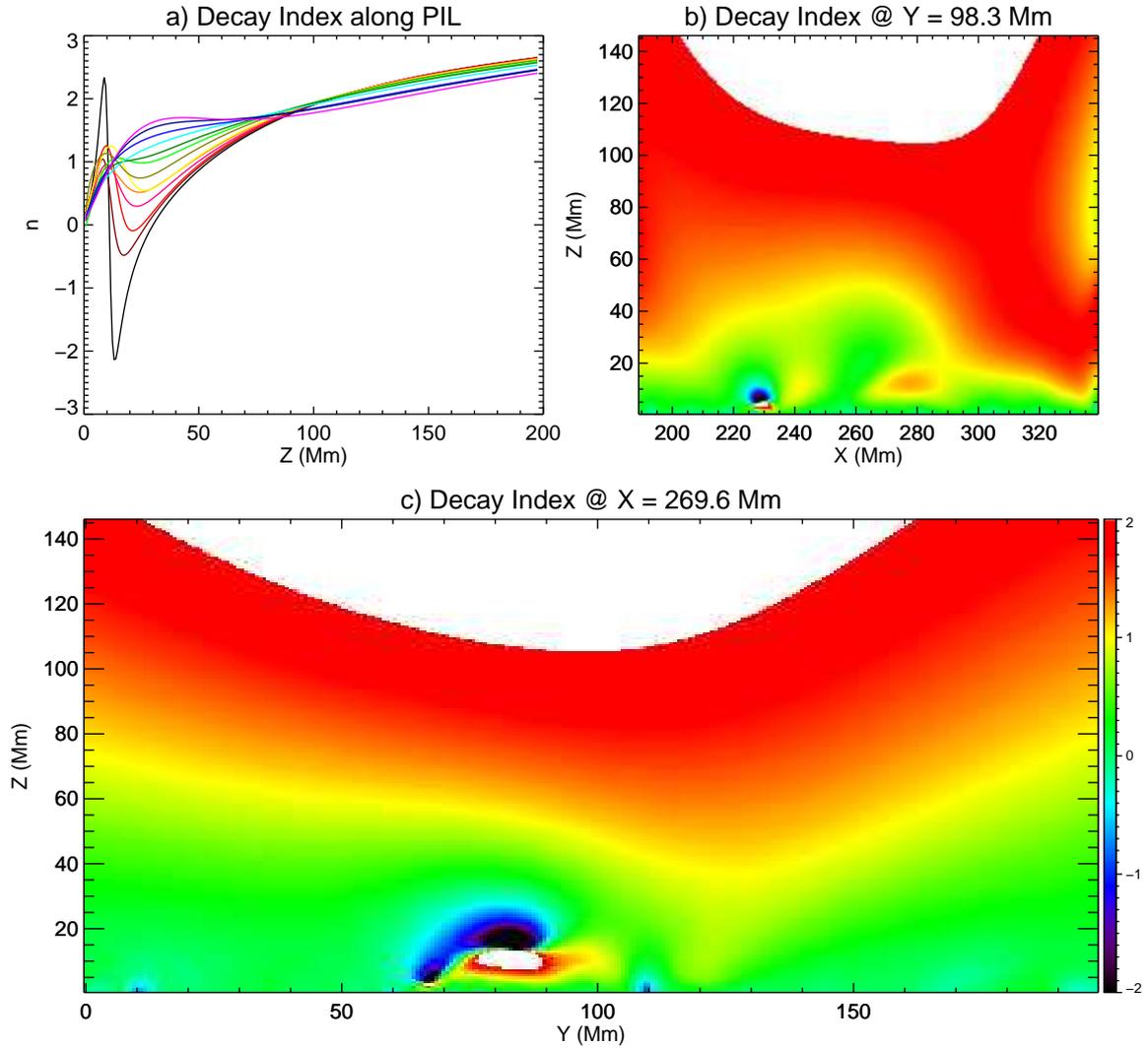}
	\caption{Decay index of the coronal potential field. a) Decay index as a function of $Z$ at the selected points (colored stars in Figure~\ref{fig:topology}) along the major PIL. Panels (b) and (c) show decay index in the $XZ$ and $YZ$ cutting planes, respectively, saturated at 2 in white and -2 in black (see the color bar). The positions of the cutting planes in the $XY$ plane are indicated by dashed lines in Figure~\ref{fig:topology}.\label{fig:decay}} 
\end{figure}

Further, the DQSLs may be conducive to the confinement of the filament, albeit a partial one. On the one hand, the two major flare ribbons in the chromosphere, i.e., the footpoints of the post-flare arcade, match well with the high-Q lines on the surface (Figure~\ref{fig:topology}). In particular, the northern ribbon is apparently bounded by the northern section of \dqw's footprint. On the other hand, the region around the two DQSLs' intersection corresponds to strong magnetic confinement as suggested by small decay indexes ($n\le 1$; Figure~\ref{fig:decay}(b and c)) at altitudes up to as high as 50 Mm. 

\begin{addendum}
	\item [Acknowledgments] R.L. acknowledges the support by NSFC 41474151, NSFC 41774150, and the Thousand Young Talents Program of China. This work was also supported by NSFC 41421063, CAS Key Research Program of Frontier Sciences QYZDB-SSW-DQC015 and the fundamental research funds for the central universities.
\end{addendum}
\renewcommand{\refname}{References}

\clearpage

\end{document}